 \documentclass{aa}
\usepackage{epsfig,amsmath,amssymb}

\def\msol{M_\odot}

\def\te{T_{\rm eff}}
\def\vrot{{\rm v}_{rot}}
\def\vconv{{\rm v}_{conv}}
\def\tconv{{\rm t}_{conv}}
\def\hp{{H}_{P}}
\def\kms{\,{\rm km}\,{\rm s}^{-1}}
\def\cms{\,{\rm cm}\,{\rm s}^{-1}}
\def\cm2s{\,{\rm cm}^2\,{\rm s}^{-1}}
\def\dyn{\,{\rm dyn}\,{\rm cm}^{-2}}

\def\ga{\,\hbox{\hbox{$ > $}\kern -0.8em \lower 1.0ex\hbox{$\sim$}}\,}
\def\la{\,\hbox{\hbox{$ < $}\kern -0.8em \lower 1.0ex\hbox{$\sim$}}\,}

\begin{document}

\title{Evolution of low-mass star and brown dwarf eclipsing binaries}

\author{Gilles Chabrier, Jos\'e Gallardo, Isabelle Baraffe }

\institute{Ecole Normale Sup\'erieure de Lyon, CRAL (UMR CNRS 5574), Universit\'e de Lyon, France}

 \date{Received/Accepted}

\titlerunning{Evolution of low-mass star and brown dwarf eclipsing binaries}
\authorrunning{Chabrier et al. }
\abstract{
We examine the evolution of low-mass star and brown dwarf eclipsing binaries. These objects are rapid
rotators and are believed to shelter large magnetic fields.}
{ We suggest that reduced convective efficiency, due to fast rotation and large field
strengths, and/or to magnetic spot coverage of the radiating surface
 significantly affect their evolution,
leading to a reduced heat flux and thus larger radii and cooler effective temperatures than for regular objects.}{We have considered such processes in our evolutionary calculations, using a phenomenological approach.}
{This yields mass-radius and effective temperature-radius
relationships in agreement with the observations. We also reproduce the effective temperature ratio
and the radii of the two components of
the recently discovered puzzling eclipsing brown dwarf system.}
{These calculations show that fast rotation and/or magnetic activity may significantly affect the evolution of eclipsing binaries and
that the mechanical and thermal properties of these objects depart from the ones of
non-active low-mass objects. We find that, for internal field strengths compatible with the
observed surface value of a few kiloGauss, convection can be severely inhibited.
The onset of a central radiative zone for rapidly rotating active low-mass stars
might thus occur below the usual $\sim 0.35\,\msol$ limit. }


 \maketitle
 
\section{Introduction}

Low-mass stars (LMS), i.e., M-type stars in the present context, represent an overwhelming fraction of the Galactic stellar population (\cite{Chabrier03}). 
Observational determination of their 
mass-radius relationship provides a
stringent testing of the theoretical description of their structure and evolution. Although observations of this
relationship agree well with the
theoretical predictions in a large number of cases (\cite{Segransan03}; \cite{CBAH05}), the radius determinations of eclipsing binaries (EBs) depart significantly ($\sim$10-15\%)
from these predictions
(\cite{Torres}; \cite{CBAH05}).
These discrepancies are unlikely to be due to inaccurate equation of state or opacities, whose treatment
is relatively well mastered for $m\ga 0.4\msol$ and $\te\ga 4000$ K (Chabrier \& Baraffe 2000, CB00).
Furthermore, active LMS are observed to be redder, i.e. cooler, than the
other LMS (\cite{Hawley96}). These particular behaviours pose a challenge to theorists. On the other hand, the majority of  LMS, notably
EBs, are fast rotators and exhibit strong persistent
coronal (X-ray) and chromospheric (H$_\alpha$) activity or large
flares (\cite{Gizis02}), indicating the presence of a strong magnetic field (\cite{Donati06}). In this Letter, we suggest that the discrepancy between the observed mass-radius relationship of EBs
and of other LMS does not stem from inaccurate equation-of-state or opacity problems, but is due
to a rotation or magnetic field induced reduction of the efficiency of large-scale thermal convection in their interior, leading to less efficient heat transport. We show
that the reduction of the star's radiating surface due to magnetic spot coverage also yields a
smaller $\te$ and a larger radius.

\section{Effect of rotation and magnetic field}
The EBs and, more generally, active LMS are fast rotators
with rotation periods of
$P\la 3$ days and rotation velocities of $\vrot \ga 10\kms$ (\cite{Delfosse98}; \cite{Reid02,MB03}). Typical convective time scales in their interior are $\tconv$$\sim$${H_P \over \vconv}$$\sim$${10^9\over (10^2-10^3)}$$\sim$$10^6$-$10^7$ s, where $H_P$$\sim$$ R_\star$
 is the pressure scale height (\cite{CB00}). This yields Rossby numbers
$Ro$=${P\over \tconv}\la 10^{-2} $. At a very small Rossby number, the Proudman-Taylor theorem enforces fluid motions to columnar or sheet-like structures 
with a characteristic length scale perpendicular to the rotation axis that is much smaller than the one 
along the rotation axis (of the order of $R_\star$).
Although the conditions of this theorem do not exactly apply for spherical objects
with finite viscosity,
rotation-dominated convective motions are severely affected by rotation,
leading to highly anisotropic patterns with
large-scale fluid motions being confined to columns
along the rotation axis and motions in the other directions being strongly inhibited 
(\cite{ZhangJones97}).
As a whole, this reduces the characteristic length scale along these directions and the mean convective velocity amplitudes,
thus the efficiency of thermal convection to transport the internal heat flux.

On the other hand, fully ionised interiors of LMS are excellent electrical conductors.
The interaction between the magnetic field and the fluid motions induces an electrical current and a
Lorentz force, which in turn affects the motions. 
In the presence of a magnetic field,
the anisotropy of rotation-dominated convection is reduced and the large-scale flows reach the so-called
MAC (Magnetic, Archimedean, Coriolis) balance between the buoyancy, Coriolis, and Lorentz forces (\cite{StarchenkoJones}).
Equipartition
between the buoyancy force, $\sim \rho g\delta (\Delta T/T)$ (where $\delta$=$-({\partial \ln \rho \over \partial \ln T})_P$ and $\Delta T$
is the temperature excess over a pressure scale height), and the Coriolis force, $\sim 2\rho\Omega \vconv$
(where $\Omega={\vrot\over R_\star}$
is the star angular velocity), yields

\begin{eqnarray}
v_{conv} \approx {g\delta \Delta  T\over T \Omega}
 \label{vconv}
\end{eqnarray}

\noindent for the typical {\it average} large-scale velocity. For $\Omega \sim 10^{-4}$ s$^{-1}$, typical values of the
various quantities over the LMS domain give $\vconv \sim$10-100$\cms$, about
a factor of 10 lower than the predictions of the usual mixing length theory.

There is presently no clear understanding of the
interaction between convection and magnetic field under stellar interior conditions.
We can, however, estimate the conditions for generation of a dynamo
in LMS or BD interiors. The microscopic magnetic diffusivity of metallic hydrogen is $\eta \approx 10^2$-$10^3$ $\cm2s$. According to Ohm's law and Maxwell's equations (hydromagnetic induction equation), a magnetic field will decay unless a velocity field can counteract or balance the diffusive effects.
For typical values of the convective velocity (see above), the characteristic magnetic Reynolds
number over a star-size conducting region is
$R_m=v_{conv} \,R_\star/\eta \gg 100$. According to dynamo theory, $R_m$ in LMS and BD interiors is
thus large enough for dynamo to occur, providing both rotation and convection are present.
Once the criterion for dynamo onset is satisfied, the field will grow and is supposed to equilibrate when
the Lorentz and the Coriolis forces become comparable (Elsasser number of order unity), reaching the aforementioned MAC balance. This yields an amplitude for the internal field,
$B_{eq} \approx (8\pi \bar{\rho} \eta \Omega)^{1/2}\ga 10\,{\rm G}$, for fast rotators ($\vrot \ga 10\kms$).

At large $R_m$, i.e. in the dissipationless regime, however,
 magnetic diffusion is mainly due to turbulent rather than
molecular diffusion, with $\eta \equiv \eta_t\sim l\,v_{conv}$\footnote{Observations
of sunspot decay indeed suggest that the solar surface diffusivity is many
orders of magnitude higher than the atomic value.}.
In that case,
non-linear saturation occurs when turbulence, enhancing the diffusive processes, is strong enough to reduce $R_m$ down to the critical value for dynamo action, $R_{m}\ga  50$. This yields typical magnetic length scales $l\la R_\star/50$, and thus $B_{eq}\approx {\rm a\, few}\,10^4$ G for LMS average conditions,
in good agreement with {\it surface} field determinations of a few kG (\cite{Donati06}; \cite{ReinersBasri}).
This corresponds to Alfven velocities $v_A={B\over \sqrt{4\pi \rho}} \ga \vconv$, 
so that for such strong fields the Lorentz force will impede the convection by reducing the flow speed. 
The main effect of a strong magnetic field is to inhibit motions across it in
comparison with motions along it. 
An ideally conducting fluid is tied to the fluid lines. 
In a fluid of
finite conductivity, motion across the field is possible at a rate governed by the conductivity.
In a highly conductive medium with a strong magnetic field, the motion will be slow. Stevenson's (1979)
stability analysis shows that the combination of fast rotation ($Ro\la 0.1$) and a magnetic field
with finite diffusivity {\it enhances} convection, because of the reduction by the Lorentz force of the flow anisotropy
due to the Proudman-Taylor constraint. Stevenson's approach, however, applies to planar
geometry, i.e. thin convection zones, and to uniform density and magnetic fields and is likely to break down for large (star-size) convective zones and strong fields. Magneto-convection 3D simulations indeed show that the magnetic field inhibits the magnitude of the velocity fluctuations and reduces the heat flux (\cite{Stein92}).
On the basis of the aforementioned
field strength values, it seems unavoidable to suppose that, even though
the magnetic field will not necessarily stabilise
the fluid against convection in a fluid of finite electrical resistivity, it will
cause a serious reduction of convective efficiency.

These estimates, on the other hand, show that the
magnetic pressure in LMS or BD interiors, ${B^2\over 8\pi}\la 10^7\dyn$, is orders of magnitude lower than the gas pressure, $P_{g}\ga 10^{12}$-$10^{16}\dyn$ (\cite{CB00}), and it can be safely ignored in the internal structure equations.
Indeed, if the fluid is convectively unstable, the ratio of the magnetic pressure inside the flux tube over the surrounding mean gas pressure is expected to be of the order of the superadiabaticity, 
${(B^2/8\pi)\over P_{g}}\la (\nabla - \nabla_{ad})\la 10^{-7}$ (\cite{GoughTayler66}; \cite{Meyer}).
An estimate for complete inhibition of convection is obtained when the Lorentz force is strong enough to balance the buoyancy force:

\begin{eqnarray}
{B^2\over 4\pi l}\ga \rho g \delta (\nabla - \nabla_{ad})\, ,
\label{crit}
\end{eqnarray}

\noindent where $l\ll R_\star$ is the aforementioned characteristic magnetic length scale in a dynamo
turbulent medium. This yields field amplitudes of the order of $10^4$ G, comparable to the value of $B_{eq}$. Magnetic fields can thus in principle severely inhibit convection in the interior of
active LMS and BDs. 
  The criterion (\ref{crit}) for
stability against convection in the presence of a magnetic field is similar to the one derived by
Stevenson in the dissipationless regime and by
Gough \& Tayler (their Eq.(1.2)),
except for the reducing
factor $\sim l/R_\star$ for the gas pressure term (assuming $\nabla P_g\sim P_g/R_\star$).
The Gough and Tayler approach, however, is primarily devoted to the study of surface spots, where
the magnetic and gas pressures are comparable. 
Applying this criterion to LMS interiors yields
 field strengths $\ga 10^7$ G for the inhibition of convection in the core of a 0.3 $\msol$ star (\cite{Mullan}).
It seems rather difficult to generate such strong fields. Together, near-equipartition (within a factor $\sim$10) between turbulent and
magnetic energy, $B^2/8\pi \sim \rho v^2_{conv}$, and
the fact that
$L(r)\propto 4\pi r^2(\rho \vconv^3)$ is a slowly varying quantity, yield
an amplification factor $\sim$10-100 from the surface to the central regions. 
A full field of several megaGauss would thus be in super-equipartition
and, if confined to the interior, would be unstable (Elsasser numbers $\gg 1$) (\cite{Tayler73}; \cite{MarkeyTayler73}).
Therefore, it does not seem realistic
to apply the Gough-Tayler criterion to the entire stellar structure, in particular for uniformly dense objects like LMS and BDs.

A real picture of a magnetised, convectively unstable medium is cooling flows along the magnetic flux tubes passing through the convective medium.
Part of the thermal flux, confined to the tubes, is thus carried by diffusion. This contraint on the flow patterns leads to a substantial reduction in the transport of energy and reduces the maximum possible heat flux.
Given the absence of a proper treatment of heat transport in a magnetised medium for LMS conditions, which requires multidimensional simulations over a characteristic convective length scale,
and given the exploratory nature of this Letter, we decided to stick to a minimalist approach, focusing on the reduced efficiency of global thermal
convection due to the presence of strong rotation and/or magnetic fields in EBs.
Such an approach is based on the phenomenological representation of this network of diffusive tubes surrounded by impeded field-free convective regions as a global convective system - since the large Reynolds numbers illustrate the overwhelming importance
of macroscopic motions over microscopic diffusive processes - with reduced efficiency. In the framework of the standard MLT
formalism, this translates into a mixing length parameter $\alpha =l/\hp < 1$.

\section{Effect of spot coverage}

The strong chromospheric and coronal emission in LMS of spectral types $\ga $M2-M3, 
with $L_{{\rm H}_\alpha}$ and $L_X$ about 100 times
the solar value (\cite{Gizis02}; \cite{MB03}), can be associated with an average large fraction of the radiating surface being covered with magnetic
spots. Cool spots are the illustration of the inhibition of energy transport by convective motions
(buoyancy) in a (rotating) highly conducting medium, but they also illustrate the fact that the field is not strong enough to suppress convection completely over the entire structure. This brings some justification to our admitedly simplistic approach where the aim is to explore the effect of reduced convective efficiency in EBs.
We denote $\beta=S_s/S_\star$ as the (time-averaged) fraction of the stellar surface
covered by cool spots,
${\mathcal F_s}$ the {\it total} flux emerging from the spots, and ${\mathcal F_\star}=\sigma {\te}^4_\star$ the one associated with a spot-free star of effective temperature $\te$.
Cool spots imply ${\mathcal F_s} < \beta {\mathcal F_\star}$. The total flux is

\begin{eqnarray}
{\mathcal F}&=&(1- \beta)  {\mathcal F_\star} + {\mathcal F_s}\, < \,{\mathcal F_\star}.
\end{eqnarray}

\noindent Cool spot coverage
thus yields an effective temperature $\te  <{\te}_\star $,
a consequence of the reduced heat flow reaching the surface (\cite{Cowling76}; \cite{Stein92}).
Given the presently undetermined spot temperature for LMS and the exploratory nature of the present calculations, we simply assumed black (zero-temperature) spots. The effect of spot coverage for a given value of $\beta$ is thus an upper limit.

\section{Results}

  \begin{figure}
   \centering
  \includegraphics[width=8cm]{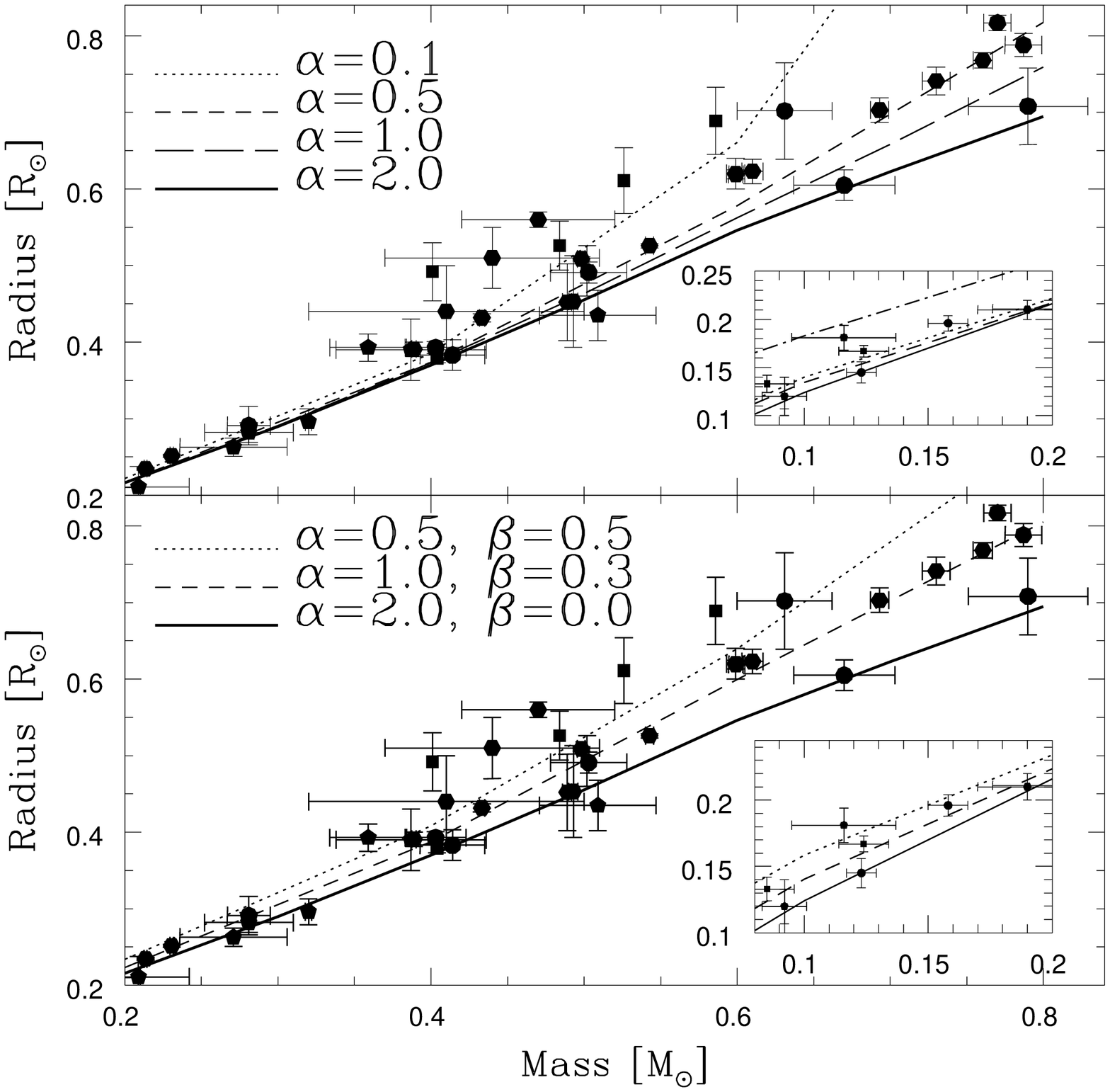}
      \caption{Mass-radius relationships for 1 Gyr old LMS for various values of the mixing length parameter
$\alpha$=$l/\hp$ ((a) upper panel) and of the fractional surface area $\beta$ covered with cool spots
((b) lower panel). Inset upper panel: dash-dot line : $\alpha=2$, $10^8$ yr. Solid thick line: standard case, no spot coverage.
Low-mass eclipsing binary observational determinations are displayed with their $1\sigma$ error bars. 
              }
         \label{fig-mR}
   \end{figure}

\subsection{Effect of reduced convective efficiency}

Figure \ref{fig-mR}(a) portrays the $m$-$R$ relationship for LMS at 1 Gyr\footnote{Objects with
$m\ge 0.09\msol$ are on the ZAMS at this age (\cite{CB00}).} for various values of
the mixing length parameter 
$\alpha$. Observed radii of EBs are displayed as well. Objects below $\sim 0.35 \msol$ are fully convective (\cite{CB97}); convection is nearly adiabatic, so changing $\alpha$ for these objects has a modest impact.
Since superadiabaticity increases with mass, in particular in the outermost regions
of the star, reducing the convective efficiency, i.e. decreasing $\alpha$, for higher masses leads to an
increasingly greater effect. The reduced convective flux requires a larger fraction of the heat to
be transported by radiation in the outermost regions, yielding a steeper outer thermal gradient
and thus a cooler $\te$. The immediate consequence is to decrease the luminosity and thus the
central temperatures, i.e. the nuclear energy
production needed to maintain thermal equilibrium. This in turn yields
an expansion of the star, i.e. a larger radius.
For BDs, the reduced heat flux implies a slower contraction rate. 

We have also explored the fate of a 0.3 $\msol$ star when drastically reducing the convective flux
in the interior ($\alpha = 0.05$) as a result of condition (\ref{crit}). The star indeed develops a stable inner radiative zone ($\nabla_{rad}<\nabla_{ad}$) over $\sim 20\%$ of its mass, yielding a
$\sim 7\%$ larger radius and a cooler $\te$.

\subsection{Effect of spot coverage}

Figure \ref{fig-mR}(b) illustrates the effect of surface spot coverage. As mentioned above,
spot coverage yields a smaller heat flux output, i.e. cooler $\te$ and thus larger radii (since $L$
is unaffected)  at a given age compared with spot-free objects. A spot
coverage fraction $\beta \sim 30$-50\%, for a value of $\alpha =1$, can by itself reproduce
most of the observed EB radii within 1$\sigma$. 

\begin{figure}
   \centering
  \includegraphics[width=6cm]{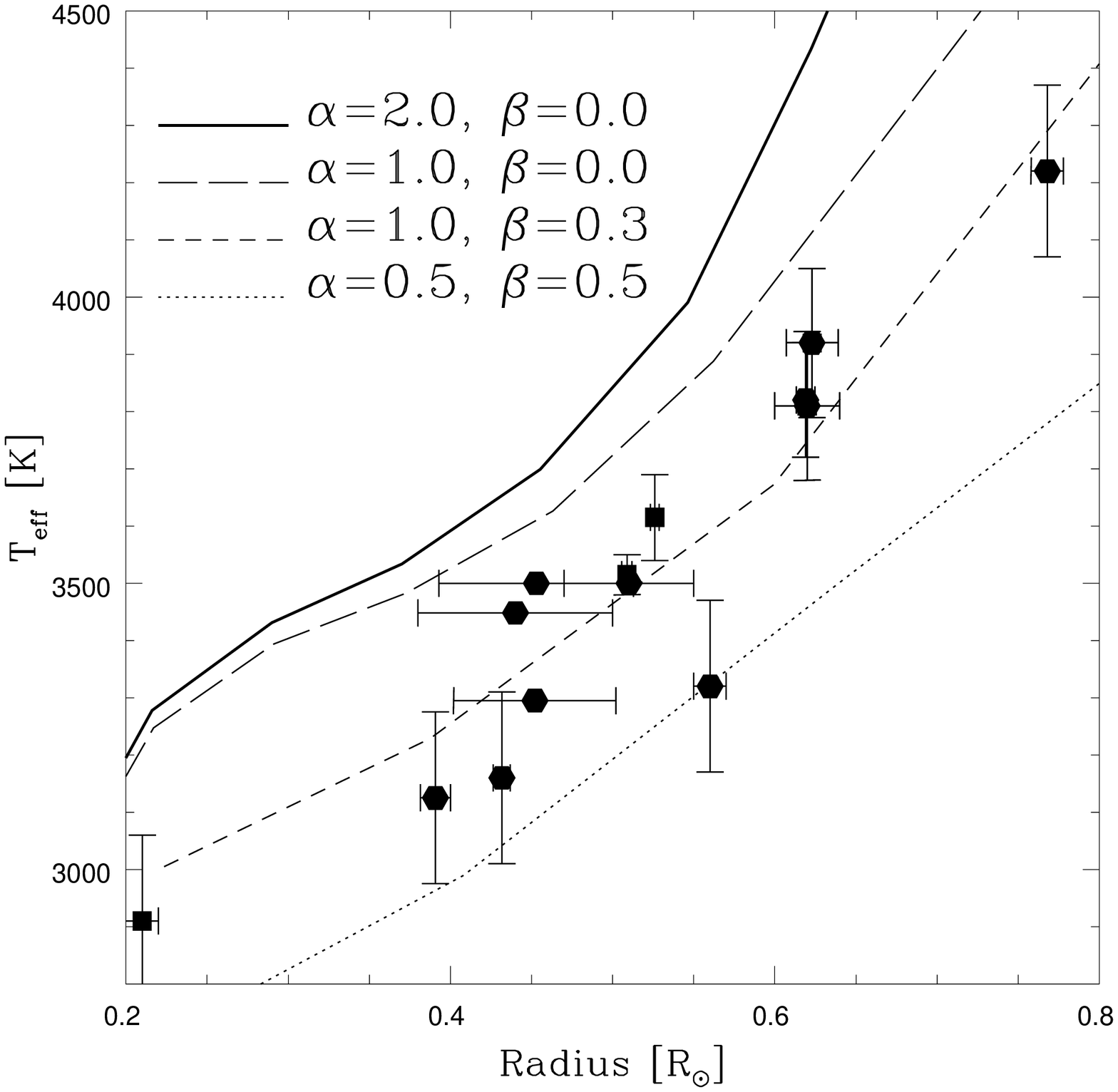}
      \caption{Effective temperature vs radius at 1 Gyr for various values of $\alpha$ and $\beta$. Solid thick line: standard case, no spot coverage. 
              }
         \label{fig-TR}
   \end{figure}

\begin{figure}
   \centering
  \includegraphics[width=6cm]{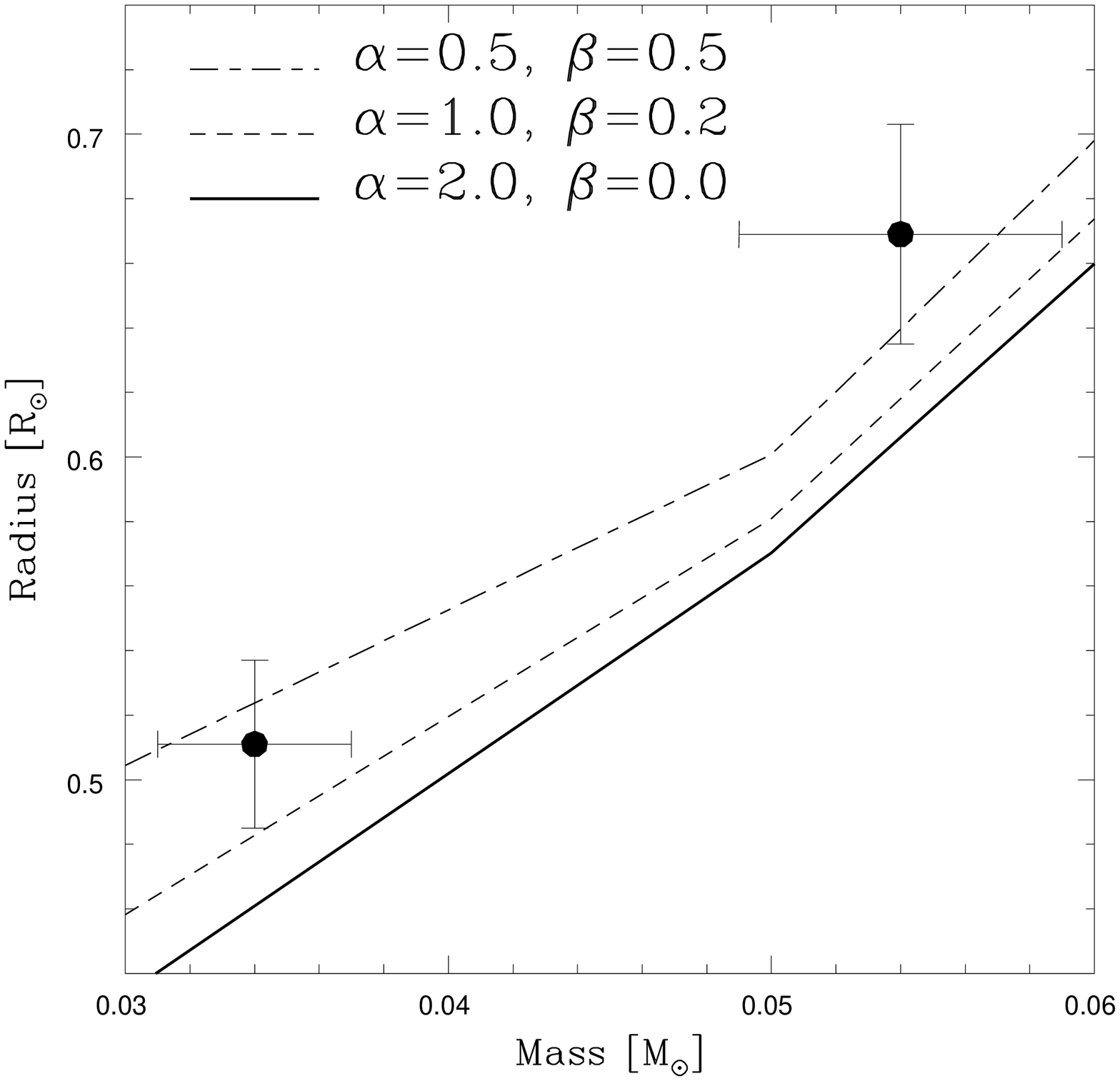}
      \caption{Same as Fig. 1 for the eclipsing BD system (Stassun et al. 2006) for different values of $\alpha$ and $\beta$.
              }
         \label{fig-BD}
   \end{figure}

Figure \ref{fig-TR} displays the effective temperature as a function of the radius for low-mass EBs for different values of
$\alpha$ and $\beta$. Both reduced convective efficiency and spot coverage yield larger radii and cooler $\te$ compared with regular objects and provide a simple explanation for
the particular mechanical and thermal properties of EBs.
Figure \ref{fig-BD} portrays $m$-$R$ relations for the recently discovered
eclipsing BD system (\cite{Stassun06}). A solution
with $\alpha=0.5$, $\beta=0.5$ for the most massive object ($m_1$=$0.054\msol$), and
$\alpha$=$1$, $\beta$=$0.2$ for the least massive one ($m_2$=$0.034\msol$) yields radius values within the error bars and a temperature {\it ratio} ${{\te}_2/  {\te}_1}$=${2440\,{\rm K}/ 2320\,{\rm K}}$=$1.05$, in agreement with the observational determination, for the age of the system $\sim 10^6$ yr\footnote{The larger $\te$ determinations obtained by Stassun et al. (2006) with the usual Sp-$\te$-colour and Sp-gravity relations suggest that these relations cannot be applied to active objects and to stellar surfaces exhibiting magnetic spots.}. This possible solution suggests that the evolution of the most massive BD has been
 significantly affected by the presence of a strong magnetic field, whereas the least massive BD has
been less affected.
Indeed, BDs with spectral type $\ga$M8-M9, i.e.
$\te \la 2400$ K, are rapid rotators and are believed to shelter large-scale magnetic fields in spite of showing no sign of persistent activity (\cite{MB03}; \cite{CK06}). 
The reason invoked for this is
the low electrical conductivity and thus the lack of substantial current generation in their atmosphere (\cite{Mohanty02}; \cite{ReinersBasri}).
Spot coverage for cool ($\te \la 2400$ K) BDs is thus expected to be small, even in the presence of a
strong magnetic field. A possible, although speculative, explanation for the
BD system of interest is that the most massive object was originally hot enough (its $\te$ for $\alpha$=$2$, $\beta$=$0$ is $\te=2800$ K) for significant spot coverage to occur, 
whereas the least massive BD remained too cool for atmospheric coupling between the magnetic field
and the gas to be significant, yielding less spot coverage.

\section{Conclusion}

In this paper, we have examined  the consequences of (i) inhibiting convection due to rotation and/or internal
magnetic field and (ii) the presence of surface magnetic spot coverage, on the evolution of LMS and BDs. We have focused on the
particular case of EBs. Our approach is phenomenological and the present paper makes no
claim to present a consistent description of the effect of a large magnetic field on the evolution of
dominantly convective objects. A proper approach to such a complex process for LMS
interior conditions requires presently unavailable
numerical tools. Our calculations show that rotation or magnetic field induced inhibition
of convection leads to a reduced heat flow and thus significantly larger radii
and cooler $\te$ than for regular objects. 
Spot coverage of the stellar radiating surface by itself has a major impact on the
evolution. Either one or a combination of these effects explains the observed mechanical and thermal
properties of EBs. Work is under way to explore the spot modulation in EBs and to derive
observational values of the spot coverage fraction $\beta$ for M-stars (Morales et al., in preparation).
A spot coverage fraction $\beta \sim 30$-50\%, however, is consistent with observations of rapidly rotating, more massive active stars (\cite{Jeffers05}).
These calculations strongly suggest that the evolution of EBs, or of very magnetically active
LMS and BDs, differs noticeably from the one of objects for which
 rotational or magnetic effects are negligible, a fact supported by observational analysis (\cite{Morales}). This
should be taken into consideration when comparing observations with theoretical models.
Global averaging of the mass-radius relationship for LMS
is thus incorrect and highly misleading.
We show that the
puzzling eclipsing brown dwarf system that was recently discovered, with the more massive companion being
the cooler one, can be explained if the evolution of the most massive
object has been strongly affected by magnetic effects.

We suggest that 
internal field strengths consistent with the observed surface value, about a few kG,
might be sufficient for severely inhibiting convection in parts of LMS interiors, pushing the limit
for the onset of an inner radiative zone
below $\sim 0.35\msol$, with important consequences for the field geometry (\cite{CK06}; \cite{Donati06}).

\clearpage

\begin{figure}
\epsfxsize=180mm
\epsfysize=200mm
\epsffile{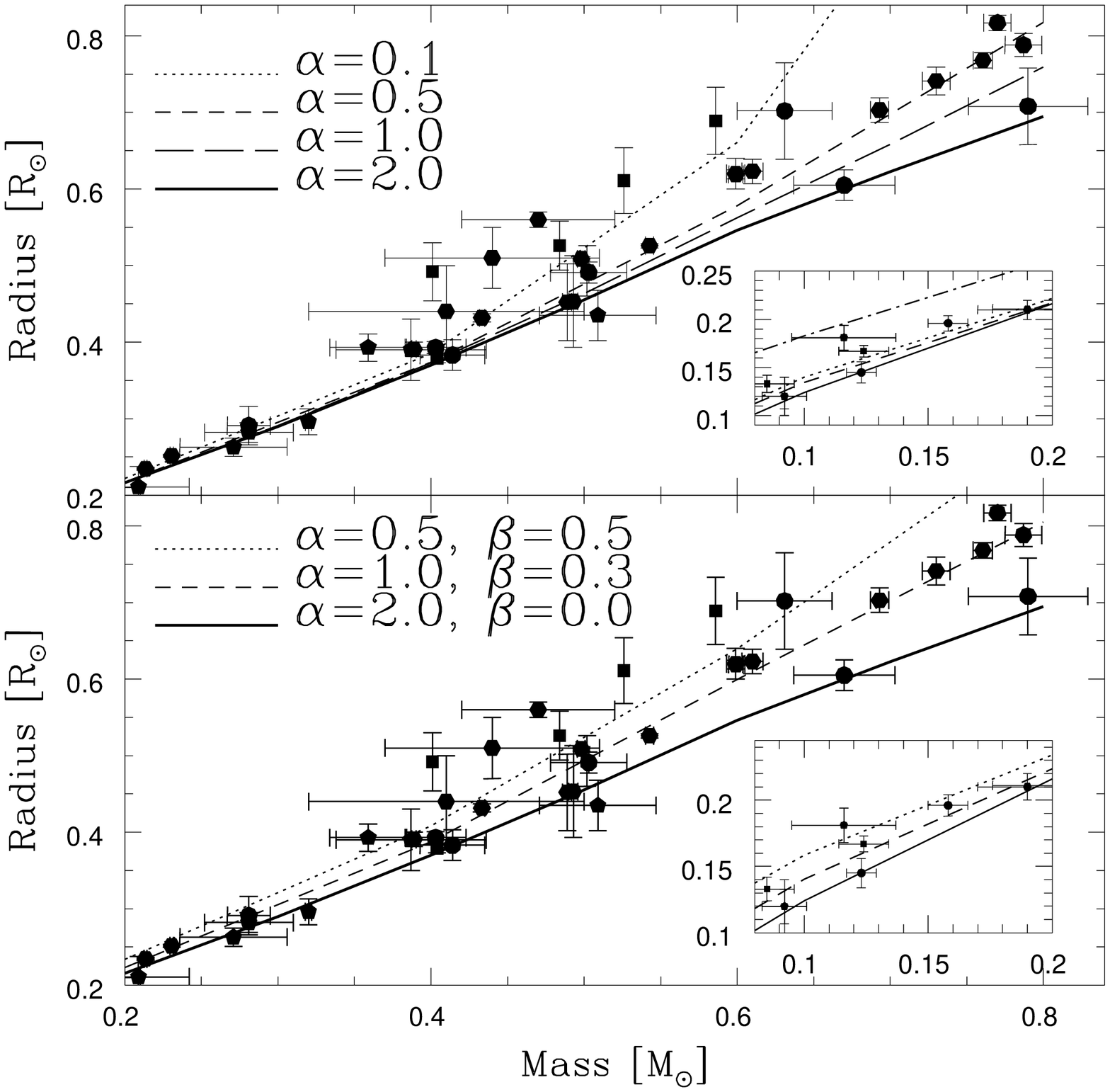}
\end{figure}

\clearpage

\begin{figure}
\epsfxsize=180mm
\epsfysize=200mm
\epsffile{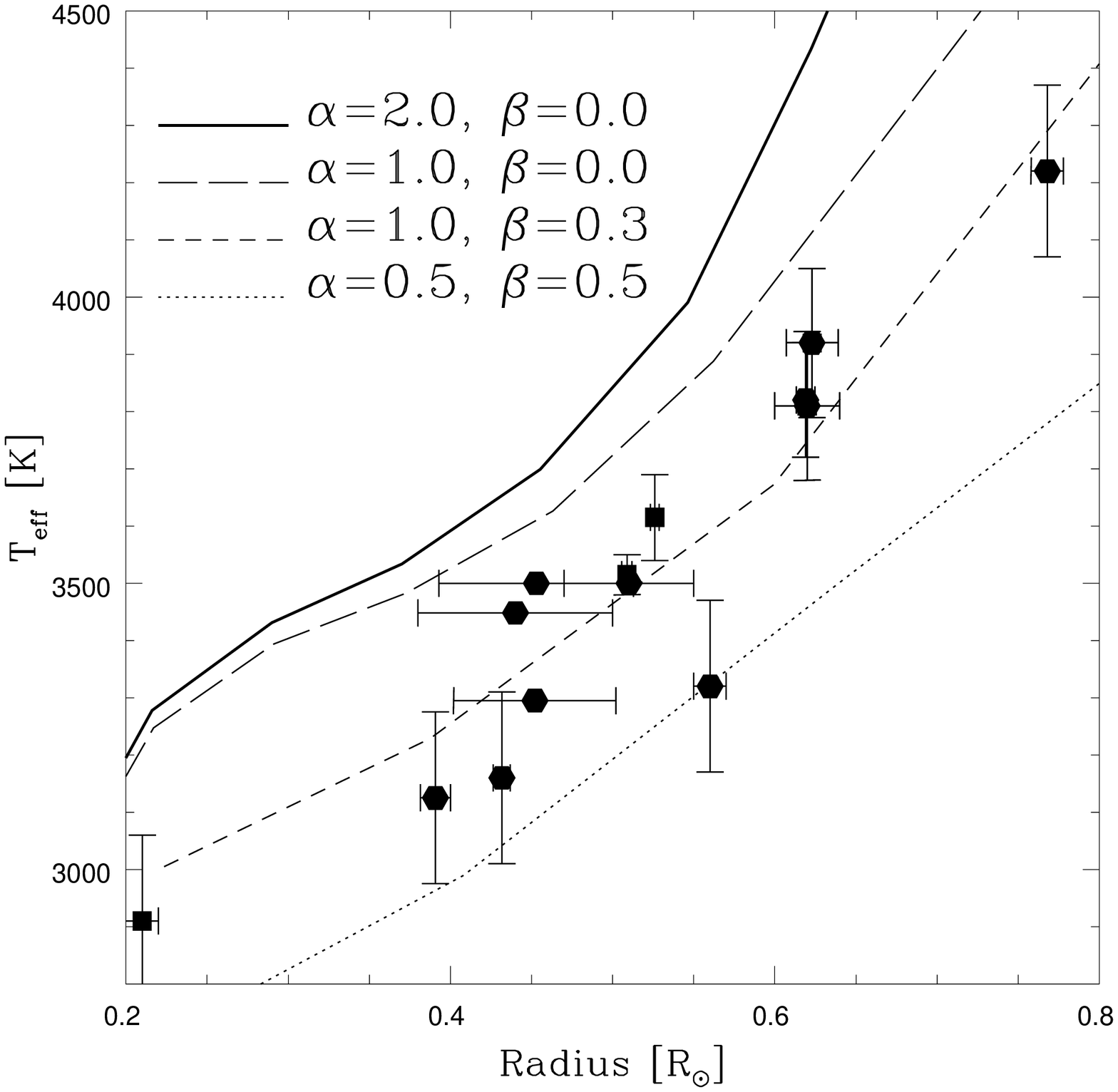}
\end{figure}

\clearpage

\begin{figure}
\epsfxsize=180mm
\epsfysize=200mm
\epsffile{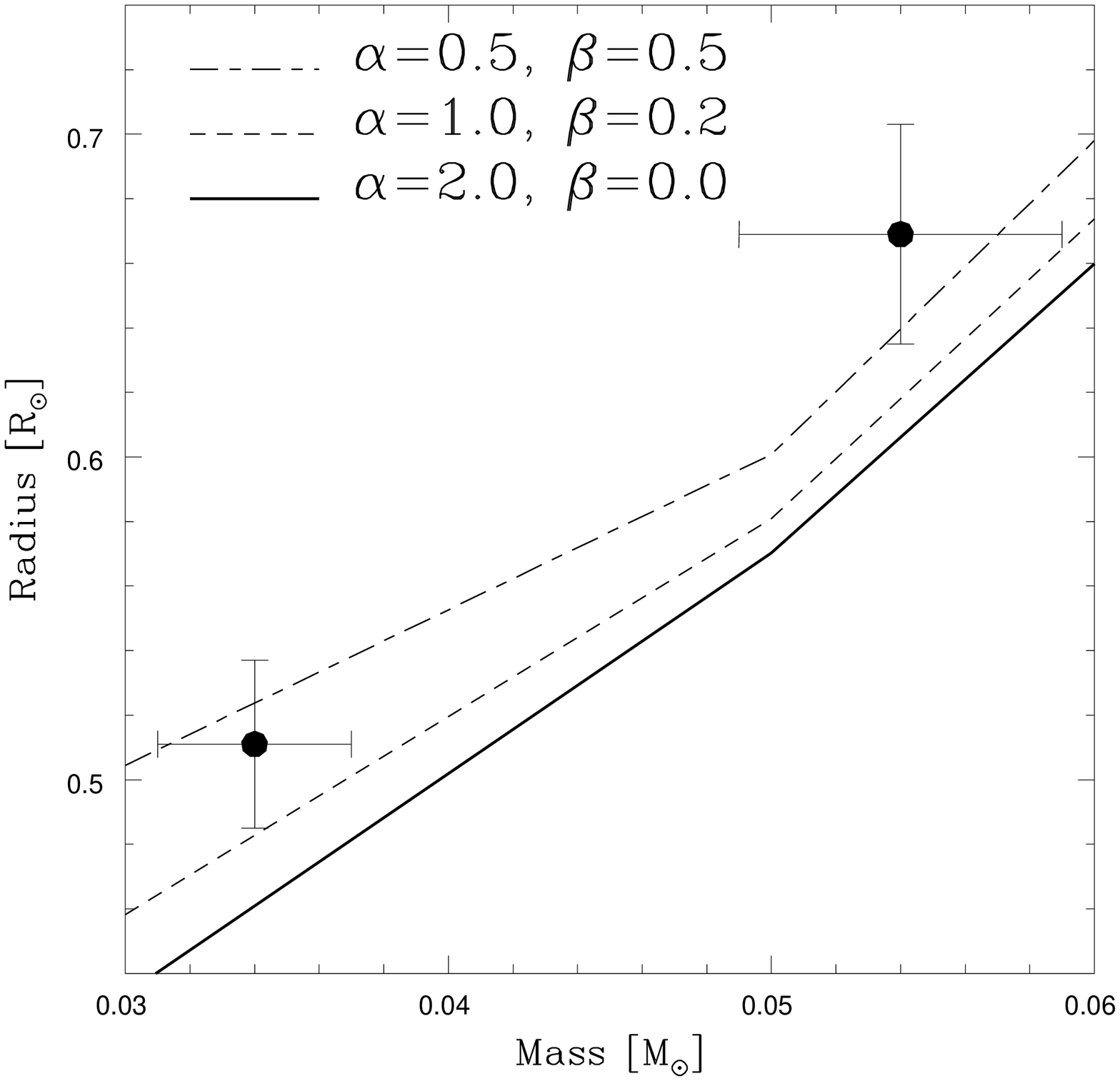}
\end{figure}

\end{document}